\documentclass[10pt, aps,  
prl,
noshowkeys,  twocolumn,
 floatfix, preprintnumbers]{revtex4-1}
\usepackage{amsmath, amssymb, amsfonts}
\usepackage{graphicx}
\usepackage{color}
\usepackage{times}
\usepackage{subcaption}
\usepackage{xcolor}

\newcommand{\bp}{{\boldsymbol{p}}}

\begin{document}
\title{Observation of Inter-layer Excitons in MoSe$_2$ Single Crystals}

\author{Jason Horng}
\affiliation{Physics  Department,  University  of  Michigan, 450  Church  Street,  Ann  Arbor,  MI  48109-2122,  USA}

\author{Tineke Stroucken}
\affiliation{Department of Physics and Material Sciences Center,
Philipps University Marburg, Renthof 5, D-35032 Marburg, Germany}

\author{Long Zhang}
\affiliation{Physics  Department,  University  of  Michigan, 450  Church  Street,  Ann  Arbor,  MI  48109-2122,  USA}

\author{Eunice Y. Paik}
\affiliation{Physics  Department,  University  of  Michigan, 450  Church  Street,  Ann  Arbor,  MI  48109-2122,  USA}

\author{Hui Deng}
\affiliation{Physics  Department,  University  of  Michigan, 450  Church  Street,  Ann  Arbor,  MI  48109-2122,  USA}

\author{Stephan W. Koch}
\affiliation{Department of Physics and Material Sciences Center,
Philipps University Marburg, Renthof 5, D-35032 Marburg, Germany}

\begin{abstract}
Interlayer excitons are observed coexisting with intralayer excitons in bi-layer, few-layer, and bulk  MoSe$_2$ single crystals by confocal reflection contrast spectroscopy. Quantitative analysis using the Dirac-Bloch-Equations provides unambiguous state assignment of all the measured resonances. The interlayer excitons in bilayer MoSe$_2$ have a large binding energy of  $153$~meV, narrow linewidth of $20$~meV, and their spectral weight is comparable to the commonly studied higher-order intralayer excitons. At the same time, the interlayer excitons are characterized by distinct transition energies and permanent dipole moments providing a promising high temperature and optically accessible platform for dipolar exciton physics. 
\end{abstract}
\date{\today}
\maketitle

The basic understanding of spatially direct semiconductor excitons dates back to the 1930s \cite{frenkel_transformation_1931,wannier_structure_1937}, where an exciton in  a single crystal has been described as a Coulomb bound pair of an electron and a hole that spatially overlap in the absence of external electrical or magnetic fields. To create spatially indirect excitons, heterostructures were used, first with coupled GaAs quantum wells \cite{Colocci_Temperature_1990, Butov_Condensation_1994} and more recently with stacked van der Waals crystal systems (vdWcs) \cite{geim_van_2013,jariwala_mixed-dimensional_2016, rivera_observation_2015,philipp_nagler_interlayer_2017,zhu_charge_2015,schaibley_directional_2016,ross_interlayer_2017}.

Here, we use confocal reflection contrast spectroscopy to show that stable, indirect exciton resonances exist in multilayer single crystals of MoSe$_2$. In these interlayer excitons, the electron and hole are confined to neighboring molecular layers and the associated optical resonances are shown to co-exist with those of the usual intralayer excitons. Mono-, bi-, tri- and few-layer up to bulk crystals on the same substrate are studied systematically and bilayer results are presented with and without encapsulation.

The experimental spectra are analyzed by numerically solving the coupled microscopic  gap and Dirac-Bloch equations \cite{Meckbach2017a,Meckbach2017b}. For all structures, the agreement between the theory and experiment is fully quantitative and allows for an unambiguous state assignment.

The identification of interlayer excitons enriches our basic understanding of the optical and electrical properties of van der Waals crystals. Furthermore,
these interlayer excitons promise a new platform for non-equilibrium many-body physics. Similar to indirect excitons in heterostructures, they have a permanent, aligned dipole moment that leads to long-range dipole-dipole interactions and a wide range of associated quantum many-body phenomena \cite{fogler_high-temperature_2014,e.v._calman_indirect_nodate,mathieu_alloing_evidence_2014,monique_combescot_boseeinstein_2017,su_spatially_2017,berman_superfluidity_2017}.

Due to their permanent dipole moment, indirect excitons in conventional semiconductor heterostructures are sensitive to local electronic disorders including those at the internal interfaces of the heterostructures. They also have a greatly reduced binding energy and optical transition rate, limiting their operation temperatures and experimental accessibility. In contrast, interlayer excitons in monocrystalline vdWCs do not experience internal interfaces. Furthermore, the external crystal surfaces can be very effectively passivated by hexagonal-boron-nitride (hBN) layers. These near ideal structural conditions reflect themselves in the measured narrow linewidth of 19~meV, which is comparable to that of intra-layer excitons and many times narrower than indirect excitons in vdWc heterostructures. Moreover, due to the close proximity of the neighboring molecular layers, the interlayer excitons maintain a relatively large exciton binding energy, 153~meV in MoSe$_2$. Therefore, they may allow dipolar exciton studies at the presence of relatively high doping densities and temperatures. The interlayer excitons also remain optically active, which will enable convenient optical access to the system and potentially powerful cavity effects \cite{deng_exciton-polariton_2010,carusotto_quantum_2013}. While it will also result in a relatively fast radiative decay and restricts the system to non-equilibrium regime,  quantum many-body phenomena have been shown to survive in non-equilibrium systems.
Lastly, bi-layer and few-layer vdWcs are easier to fabricate than heterostructures and do not suffer from lattice mismatch or angle rotation between the constituting lattices.

\begin{figure}[b]
\includegraphics[width =\columnwidth]{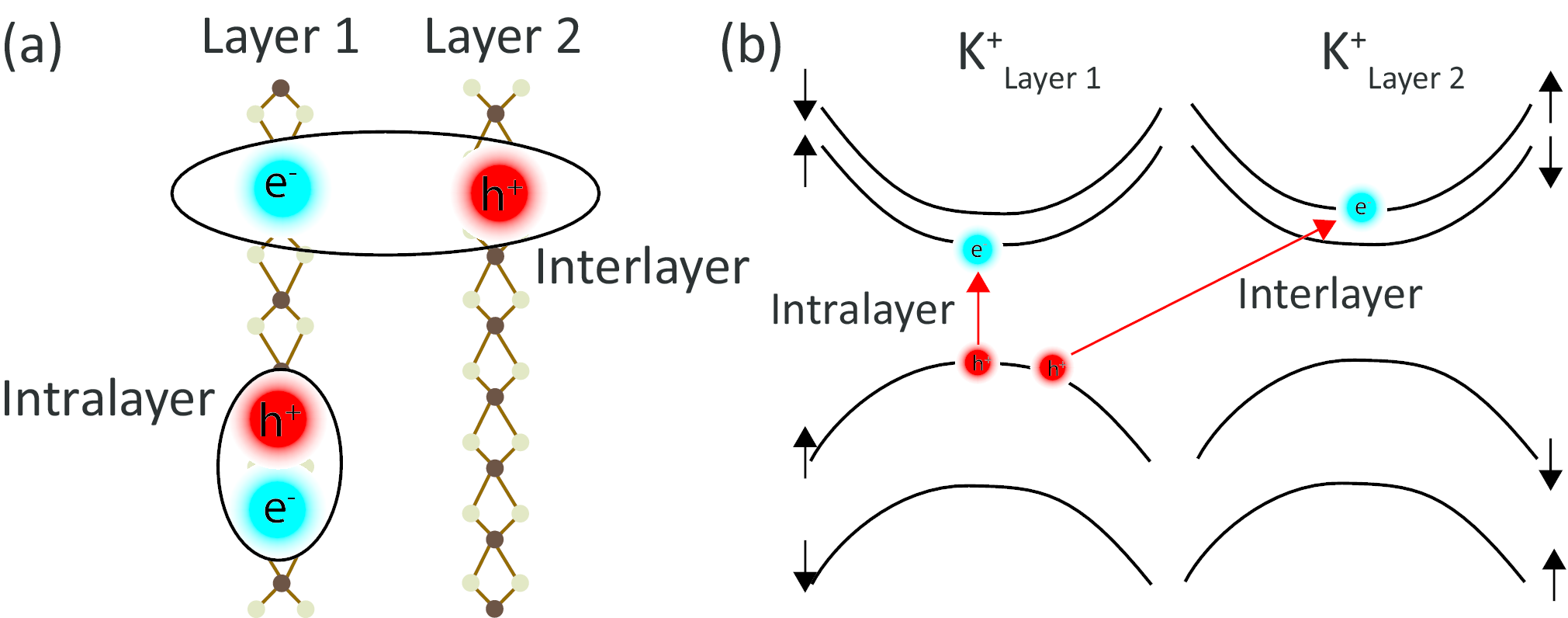}
\caption{\label{schematics} Schematic illustration of inter- and intralayer excitons in transition-metal-dichalcogenides(TMD) multilayers. (a) real-space representation of different species of excitons. Interlayer excitons consist of an electron and a hole in different layers, while the intralayer exciton consist of an electron and a hole in the same layer. (b) k-space and spin configuration for optically bright interlayer and intralayer excitons. The arrows indicate the dipole allowed transitions at the K-point of the joint Brillouin zone corresponding to the A-exciton series.  At the $K^+$ points, the layer indices are reversed.
}
\end{figure}

\begin{figure*}[t]
\includegraphics[width =\textwidth]{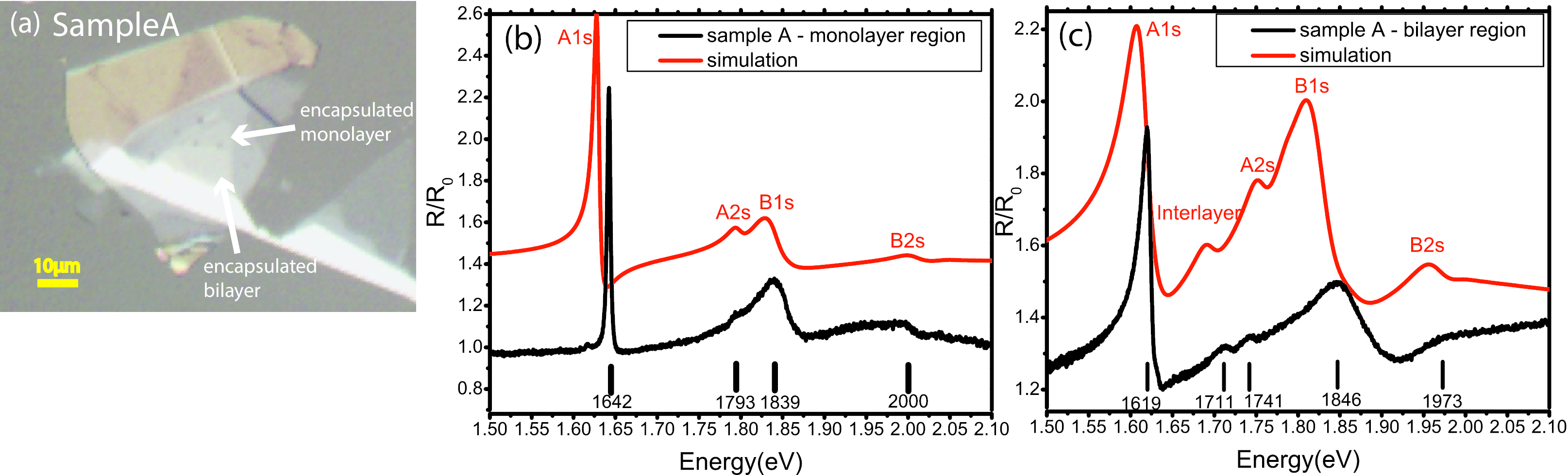}

\caption{\label{Figure2} (a) Optical microscope image of sample A, consisting of hexagonal boron nitride(hBN) encapsulated MoSe$_2$ layers. The red and black dash line indicate the encapsulated monolayer and bilayer region. (b) and (c) Measured (black curve) and simulated (red curve) reflection contrast spectra($R/R_{0}(E)$) from encapsulated monolayer(in (b)) and bilayer(in (c)) region of sample A, where $R(E)$ and $R_{0}(E)$ are reflection spectrum from sample and substrate area. The experimentally observed peaks are labeled with their resonance energies. Simulated optical spectra are based on DBE and a four-band effective Hamiltonian. Comparing experiment and theory, we assign the peaks from low to high energy as $A_{1s}$, $A_{2s}$, $B_{1s}$, $B_{2s}$  for the intra-layer excitons and identify the new peak
at 1711~meV in the bilayer as the inter-layer $A_{1s}^{\rm inter}$ exciton. All simulated spectra have been shifted for better visibility.
}
\end{figure*}

In Fig. \ref{schematics}(a), we schematically show  the real-space configuration of the interlayer exciton in a MoSe$_2$ bilayer. 
Even though the transition from a direct to indirect gap in most vdWc multilayer structures already occurs for the bilayer, the direct gap at the $K$-points of the Brillioun zone is preserved even in the bulk limit, with a dispersion that is flat along the $K-H$ direction\cite{ye2015}. This flat dispersion indicates that even in a multilayer structure, the near $K$-point quasi-particles can be considered as effectively two-dimensional, and are strongly confined within individual layers, with the potential to build bound interlayer excitons even in multilayer structures consisting of identical monolayers.

In Fig. \ref{schematics}(b), we show the band and spin configuration for the non-interacting band structure around the $K^+$-point in an A-B stacked MoSe$_2$ bilayer.
Whereas the $K^\pm$-points of a monolayer are nonequivalent and  related by the parity transformation, they are equivalent in an inversion symmetric bilayer. At each $K$-point, the non-interacting bandstructure is composed of the spin-split valence and conduction bands of layer 1 and a mirror identical copy with reversed spin ordering of layer 2\cite{xu_spin_2014}.  Consequently, dipole allowed intralayer excitons corresponding to the A-series  in MoSe$_2$ are formed by an electron in the lowest conduction band and a hole in the highest valence band, whereas the correspondent interlayer exciton uses the upper spin-split conduction band (see arrows in Fig. \ref{schematics}(b)). Thus, the optical selection rules for the interlayer excitons exhibit similar symmetry properties as those for the intra-layer excitons with the difference that the spin-valley selectivity in a monolayer is replaced by spin-layer selectivity for the excitation with circular polarized light.

To experimentally identify interlayer excitons in vdWc multilayers, we perform confocal reflection spectroscopy with a tungsten lamp light source at 5~K to study the bound electron-hole pairs on MoSe$_2$ flakes with a spatial resolution of $2~\mu$m. Signals from the sample were normalized against a point on nearby substrate to produce reflection contrast.
Fig. \ref{Figure2}(a) shows an optical microscope image of sample A, consisting of an hBN-encapsulated monolayer and bilayer MoSe$_2$.
The measured reflection contrast spectrum ($R/R_{0}$) for the monolayer region on this sample is shown as black curve in Fig. \ref{Figure2}(b), where $R$ and $R_{0}$ are reflection spectrum taken from sample and substrate area, respectively.
A typical spectrum of monolayer MoSe$_2$ was observed with reflection peaks corresponding to $A_{1s}$ ($1642$~meV) and $B_{1s}$ ($1839$~meV) excitons.
Due to the sharp linewidth resulting from the hBN encapsulation, we can also identify the excited excitons $A_{2s}$ at $1793$~meV and $B_{2s}$ at $2000$~meV, both with much smaller oscillator strengths relative to the $1s$ states.
Because of the rapid decrease of spectral weight with increasing quantum number and interference between different species of excitons, we cannot resolve the $3s, 4s ,\dots$ excitons states experimentally.

The corresponding optical spectrum in the encapsulated bilayer region is shown in Fig. \ref{Figure2}(c). The bilayer spectrum has somewhat broader $A_{1s}$ and $B_{1s}$ resonances and we note a red shift of the dominant $A_{1s}$ resonance of $23$~meV and a small blue shift of the $B_{1s}$ resonance of $7$~meV, respectively.  Strikingly, we also observe two additional peaks  above the $A_{1s}$ transition at $1711$~meV and $1741$~meV  with similar oscillator strength, and a weak spectral feature at $1973$~meV. Naively, one could try to assign the two peaks above the $A_{1s}$ resonance to the $A_{2s}$ and $A_{3s}$ exciton resonances that are red shifted by the presence of the second layer. However, this assignment is unreasonable due to the similar oscillator strength of the two observed peaks.

To understand the physical origin of the observed features, we employ the theoretical framework that combines an electrostatic model for the Coulomb interaction potential in an anisotropic medium, i.e. the gap equations to determine the interacting gap, and the Dirac-Bloch-Equations (DBE) to compute the linear optical response\cite{Meckbach2017a}. Within this model, the electronic and optical properties around the K-points of the multilayer structures are treated by considering the symmetry induced spin locking of the individual layers and the inter- and intralayer Coulomb interaction. Treating the Hamiltonian of the isolated monolayer within an effective four-band model\cite{xiao2012}, screening of the bands under consideration is included dynamically, whereas screening of all remaining bands and the dielectric environment is contained in the Coulomb matrix element. The material parameters used are listed in Ref. \cite{note}.
This model is based on the observation that the direct gap at the K-points, which contributes dominantly to the optical absorption, is preserved while increasing the number of layers from a monolayer to bulk \cite{ye2015}. At the K points, the out-of-plane effective masses of the valence and conduction bands are typically much larger than those in the in-plane directions. Consequently, the out-of-plane component of the kinetic energy can be neglected and the quasi-particles at the K-points can be considered as quasi-two dimensional, well confined within the layers. These assumptions are strongly supported by recent ARPES measurements, which have revealed the two-dimensional nature of the bands at the K point of the Brillouin zone \cite{Riley2014}

The theory predicts the resonance positions $A_{1s}=1642$~meV and $A_{2s}= 1802$~meV at zero density and temperature respectively, which shift to $A_{1s}=1629$~meV and $A_{2s}= 1796$~meV in the presence of a small carrier density of $n=1.3\cdot 10^{9}/{\rm cm}^2$. The computed optical spectrum for the encapsulated monolayer is plotted in Fig. \ref{Figure2}(b) for comparison with the experiment. In the numerical evaluations, we introduced a phenomenological linewidth with full-width-at-half-maximum(FWHM) of 4 meV for the $A_{1s}$, 30 meV for other A-exciton resonances and 40 meV for the B-series. The calculation agrees well with the experimentally observed peaks in energy as well as oscillator strengths, allowing for an unambiguous state assignment of the monolayer excitons.

Next, we apply the combined gap and DBE equations to compute the linear optical response for bilayer MoSe$_2$, using the same DFT parameters for the noninteracting band structure as for the monolayer. As long as one considers only the intralayer contributions,  the Elliot formula\cite{haugkoch2009} for an arbitrary layer within the multilayer structure is formally identical to that of a monolayer. However, quantitatively, the intralayer contributions are modified via their dependence on the detailed Coulomb matrix elements which are modified by both, the presence of the substrate and the other layers. As a result, each layer in a multilayer configuration experiences a different dielectric environment, generally leading to an intrinsic inhomogeneous broadening of the resonances due to the slightly different contributions from the individual layers.
However, for an inversion symmetric bilayer, e.g. a suspended or encapsulated system, both layers are equivalent and resonances corresponding to different intralayer excitons should be degenerate, and thus, do not give rise to {\it additional} resonances. For the encapsulated bilayer configuration at zero temperature and carrier density, we theoretically find the lowest $s$-type intra-layer resonances at $A_{1s}=1624$~meV,  $A_{2s}= 1759$~meV, and $A_{3s}=1795$~meV, with a relative oscillator strength of $1,\, 1/5.7\, 1/16.3$ respectively.
Whereas the shift of the $A_{2s}$ exciton resonance agrees with the experimental observations within an accuracy of 10\%, the resonance positions and oscillator strengths of the $2s$ and $ 3s$ intralayer states do not match the experimental data, showing that these states are not responsible for the experimentally observed features.

\begin{figure}[bt]
\begin{subfigure}{0.9\columnwidth}
  \centering
  \includegraphics[width=.7\linewidth]{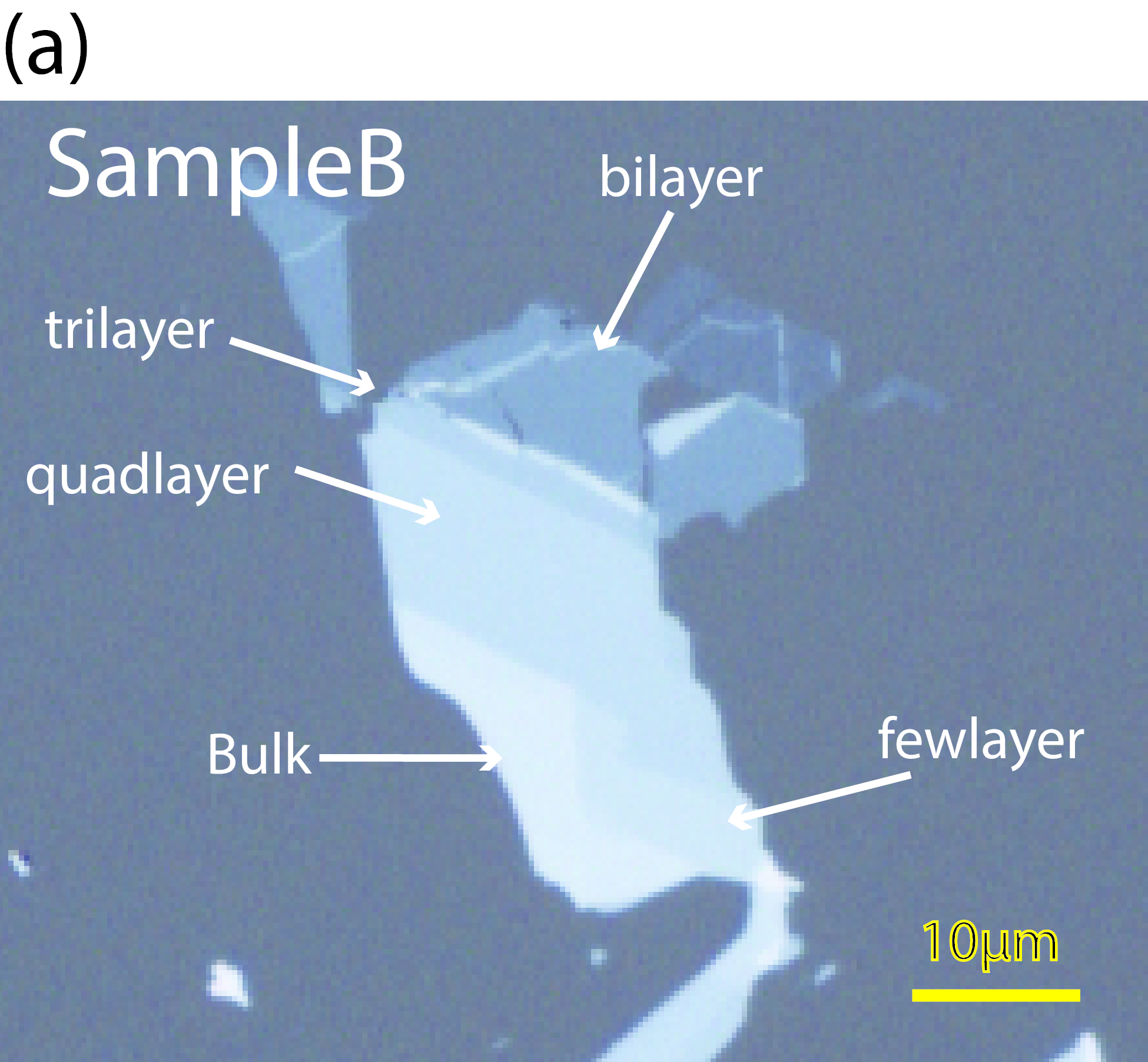}
  \label{Fig3a}
\end{subfigure}%
\\
\vspace{.2cm}
\begin{subfigure}{0.9\columnwidth}
  \centering
  \includegraphics[width=.75\linewidth]{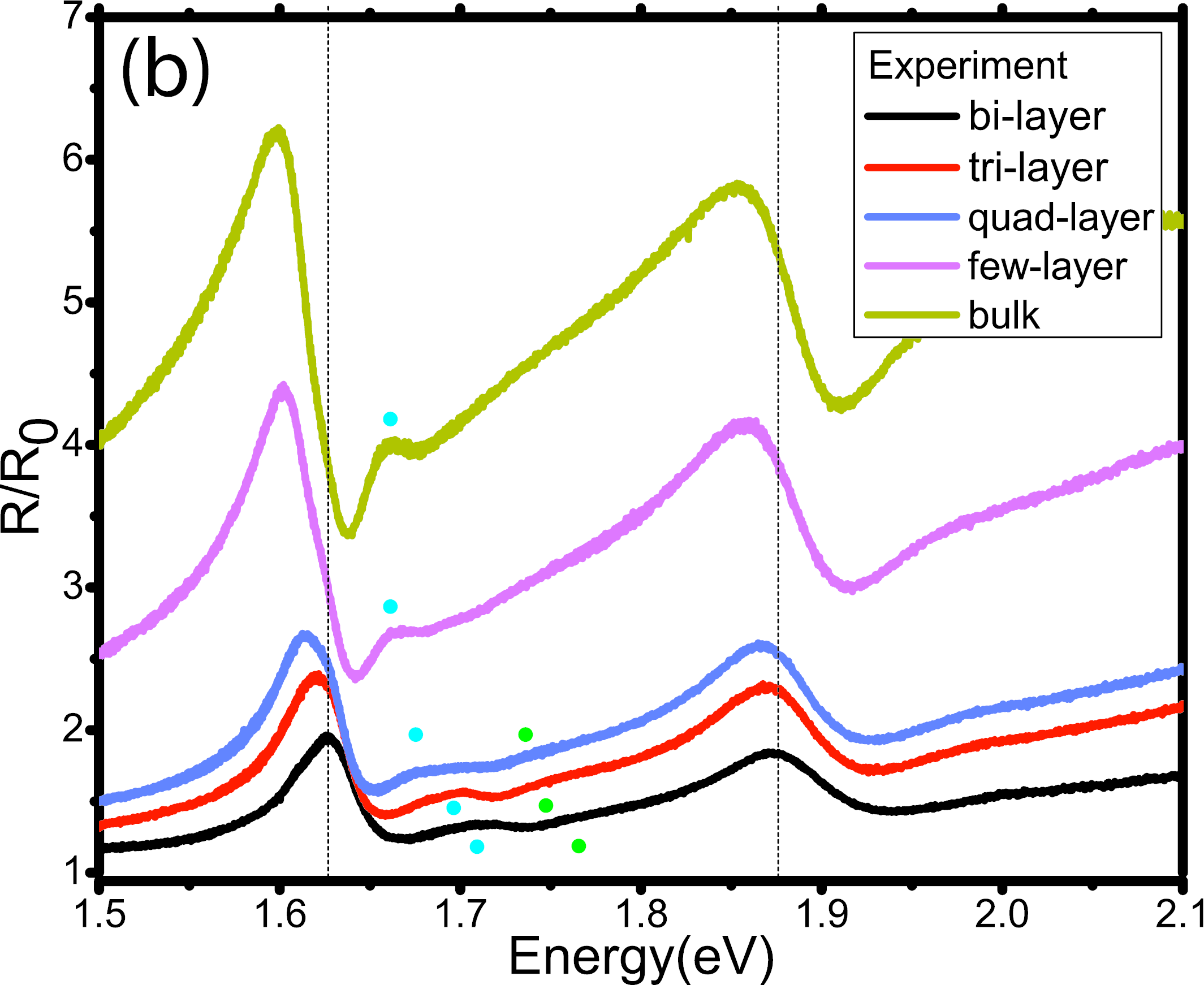}
  \label{Fig3b}
\end{subfigure}%
\\
\begin{subfigure}{0.9\columnwidth}
  \centering
  \includegraphics[width=.75\linewidth]{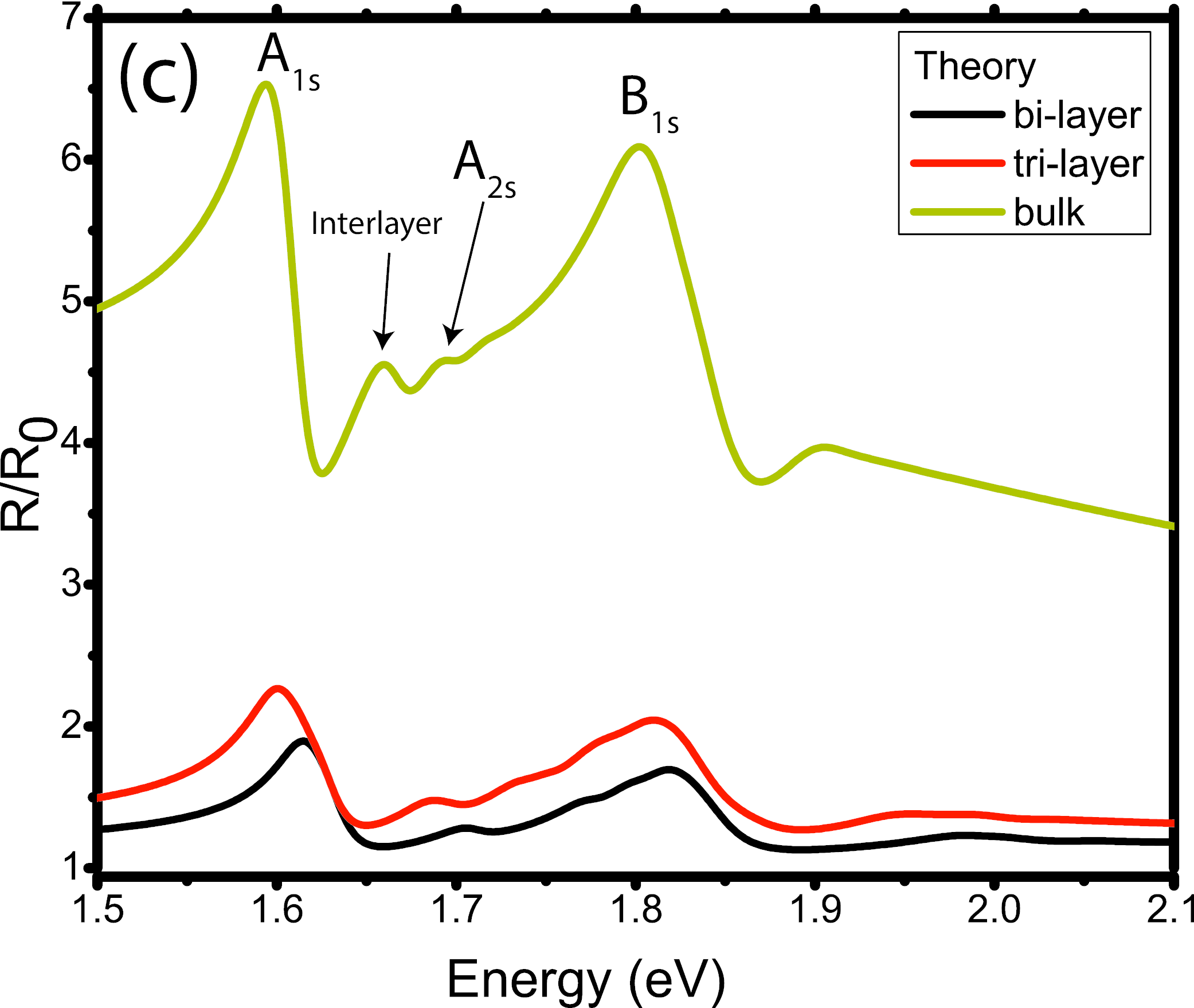}
  \label{Fig3c}
\end{subfigure}%
\caption{\label{Figure3}(a) Optical microscope image of sample B, consisting of MoSe$_2$ layers with various thicknesses on a sapphire substrate. (b) Thickness dependence of experimental linear optical spectra. The interlayer exciton resonances are shown as red arrows and show red-shift and linewidth narrow as the sample approaching bulk limit. (c) Simulated optical spectra for the corresponding sample thickness.
}
\end{figure}

In addition to the intralayer interactions, the theory predicts that the Coulomb attraction between electrons and holes
in adjacent layers should give rise to additional bound interlayer excitons. For the encapsulated bilayer configuration, we find a binding energy of
$E_{{\rm A}1s}^{\rm inter}=153$~meV for the interlayer exciton, only 27\%  less than the binding energy of $E_{{\rm A}1s}^{\rm intra}=209$~meV for the lowest intralayer exciton. As the interlayer exciton uses conduction-band states from the opposite K-valley, introducing a spin splitting of the conduction band leads to a blue shift of the A-interlayer exciton, and a simultaneous red shift of the  B-intralayer exciton. Using a $20$~meV spin splitting of the conduction band, we find the lowest inter-layer exciton at $E_{1s}^{\rm inter}=1700$~meV, between the $1s$ and $2s$ resonances of the intralayer exciton.
To determine the relative oscillator strength, we estimate the dipole matrix element for the inter-layer transition from the $d$-type Mo-orbital centered around the central $z$ positions of the adjacent layers, giving $|{\bp}_{c\nu}^{n,n\pm 1}|^2/|{\bp}_{c\nu}^{n,n}|^2\approx 0.2$. The resulting simulated spectrum for the encapsulated bilayer is shown in Fig. \ref{Figure2}(c). The calculations have been performed assuming a $20$ nm hBN capping, and we included a phenomenological broadening(FWHM) of $30$~meV and $50$~meV for A- and B- series, respectively. The results show excellent agreement with the experimentally observed spectra.  With the quantitative agreement between theory and experiment, we assign the $1741$~meV and  $1711$~meV peaks as $A_{2s}$ exciton and A-interlayer exciton, respectively.

To further support our interpretation based on interlayer exciton states, we study the influence of thickness on the resonance position and oscillator strength  in MoSe$_2$ multilayer systems.  We prepare sample B (optical microscope image in Fig.~\ref{Figure3}(a), consisting of MoSe$_2$ monolayer, bilayer, tri-layer, quad-layer and multilayers on a sapphire substrate. In Fig. \ref{Figure3}(b), we show the measured reflection contrast for various sample thicknesses. The two main transitions around $1625$~meV and $1875$~meV correspond to the $A_{1s}$ and $B_{1s}$ excitons, respectively. Aside from these two features, we also observed the interlayer peak ($1711$~meV, blue dot) and $A_{2s}$ shoulder ($\sim 1772$~meV, green dot) in the bilayer spectrum. As the layer number increases, both intralayer exciton and interlayer exciton should show a redshift behavior due to increasing dielectric screening. Experimentally, the bulk intralayer species $A_{1s}$ and $B_{1s}$ redshift about $27$~meV and $15$~meV, respectively, from bilayer samples, whereas  the observed A-interlayer shows a rather strong redshift of  $47$~meV as approaching to the bulk limit.  Similar to the monolayer, the shifts result from a simultaneously redshift of the interacting gap and exciton binding energy. As can be recognized, the interlayer exciton also becomes more apparent in the bulk limit. Apart from the narrower linewidth ($35$~meV in bilayer and $21$~meV in bulk), which is expected due to reducing inhomogeneous broadening arising from the from MoSe$_2$ surfaces, the reason for this is the relative oscillator strength. Whereas intralayer contributions increase linearly with the number of layers, the number of next neighbours and therewith the interlayer contributions increase as $2(N-1)$, thus doubling the relative oscillator strength $\Gamma^{\rm inter}/\Gamma^{\rm intra}$ going from bilayer to bulk.

We also perform DBE calculations to predict the linear optical response for numbers of layers. Again, we use the same monolayer DFT parameters including the same wavefunction overlap between adjacent layers. The computed optical spectra are plotted in Fig. \ref{Figure3}(c). For the A-exciton series, we find good agreement with the experimental data  in terms of both exciton resonance energies and oscillator strengths. In particular, we observe the increasing oscillator strength of the interlayer exciton compared to the intralayer exciton.  Whereas the shifts of the B-excitons with increasing layer thickness is predicted very accurately, the theoretical results for the multilayers show a systematic red shift of the B-exciton series as compared with experiment, which is not apparent for the monolayer. This systematic discrepancy indicates that interlayer interactions might change the DFT bandstructure going from mono- to bilayer, which is not captured by our model system. Despite this shortcoming, the qualitative and quantitative agreement between theory and experiment provides strong evidence to assign the second peak to the interlayer exciton.

We note that in the ref. \cite{chernikov_exciton_2014}, the second-lowest feature in WS$_2$ reflection spectra was assigned to the $A_{2s}$ exciton. According to our simulation and experiment, the correct assignment should be A-interlayer exciton.

\begin{figure}[bt]
\begin{subfigure}{.45\columnwidth}
  \centering
  \includegraphics[width=\linewidth]{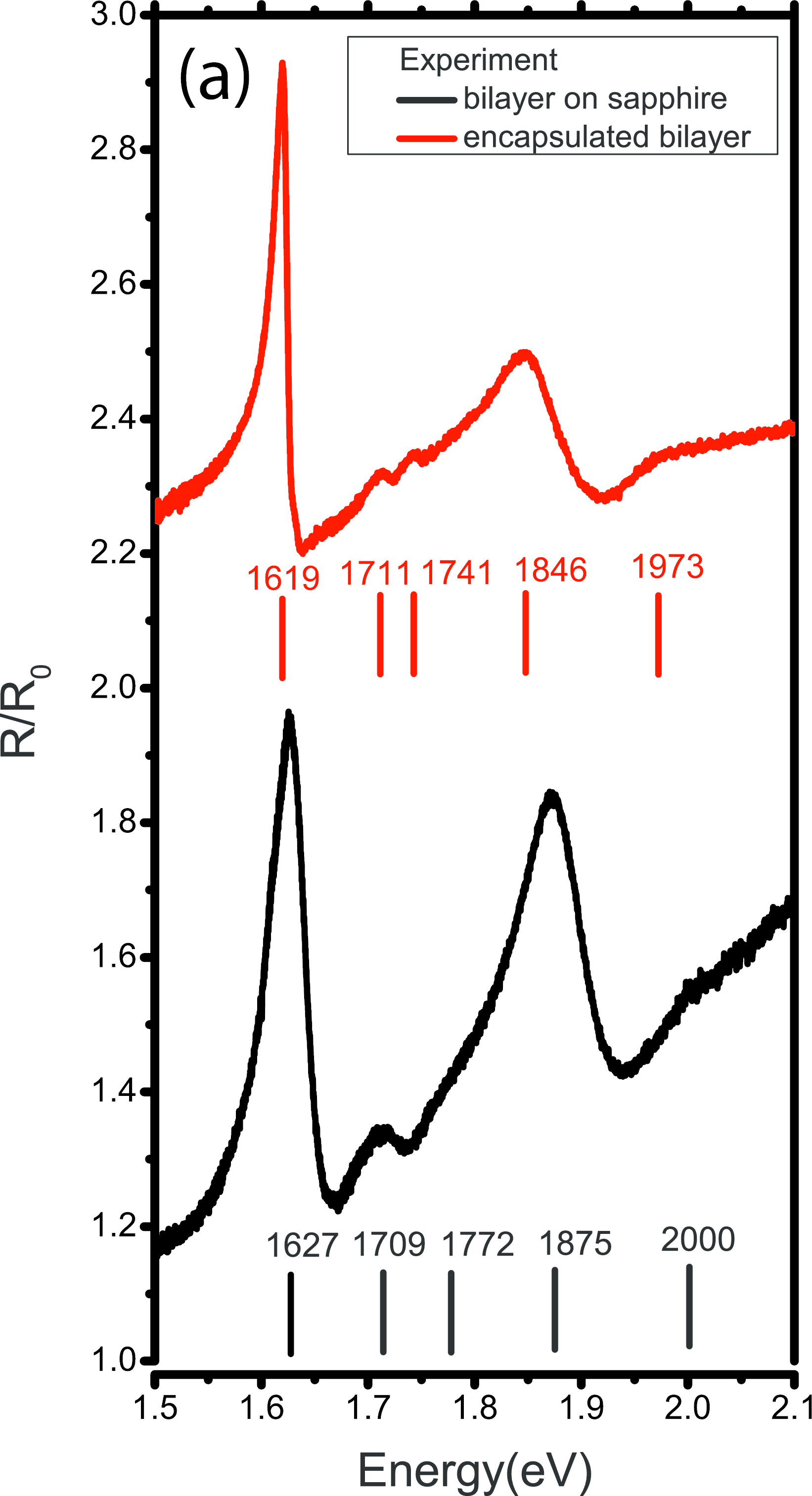}
  \label{Fig4a}
\end{subfigure}%
\begin{subfigure}{.45\columnwidth}
  \centering
  \includegraphics[width=\linewidth]{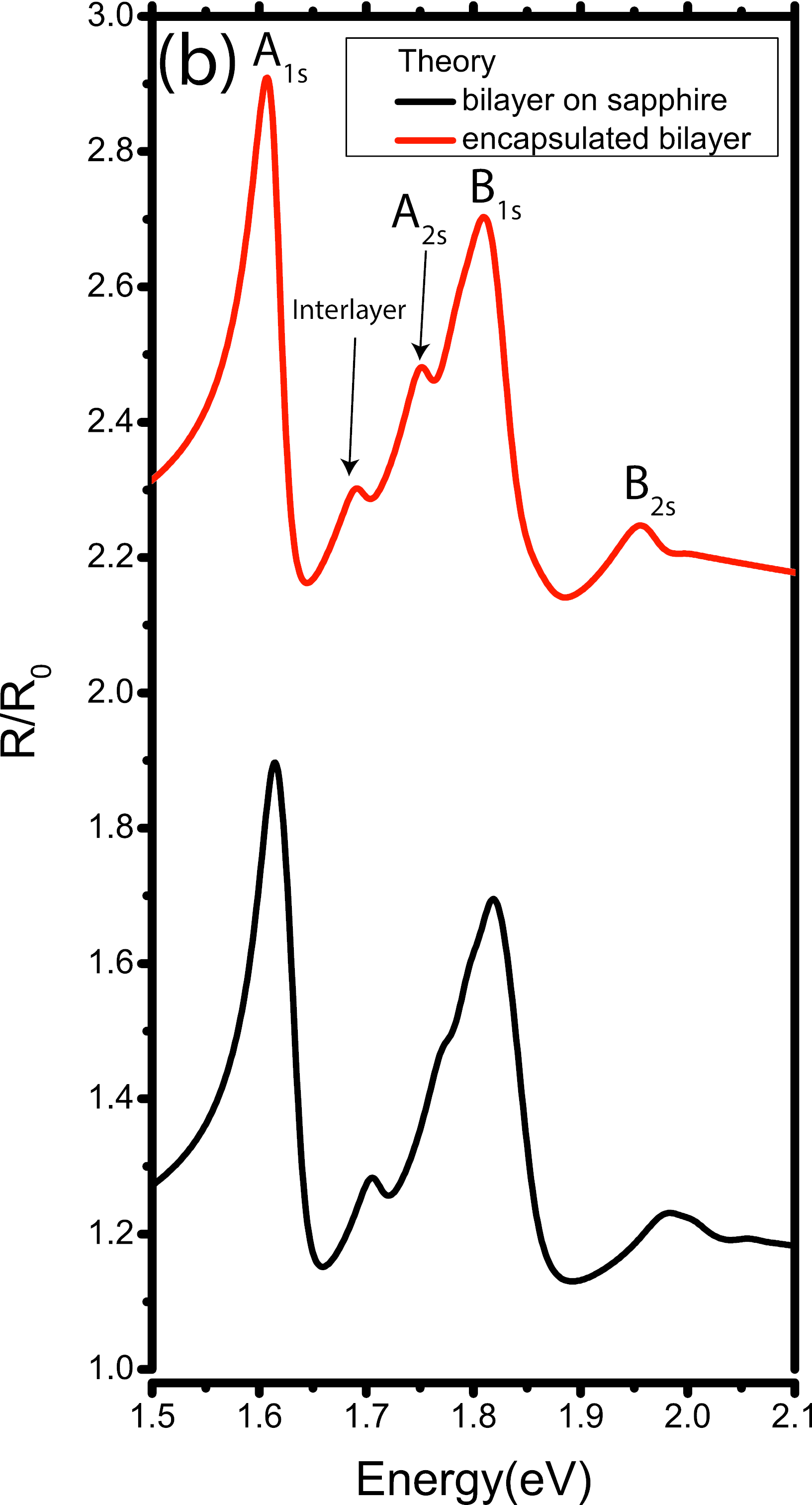}
  \label{Fig4b}
\end{subfigure}%
\caption{\label{Figure4} (a) Comparison between optical spectra of MoSe$_2$ bilayer with hBN encapsulation (red curve) and without hBN encapsulation (black curve). The MoSe$_2$ bilayer with hBN shows a significant linewidth narrowing in both intralayer and interlayer excitonic transitions possibly due to symmetric dielectric screening and a clean environment. (b) Corresponding simulated optical spectra showing the theory is able to capture the dielectric screening effects on different exciton species.
}
\end{figure}

At last, we compare the measured spectra of bilayer MoSe$_2$ with and without hBN encapsulation (Fig. \ref{Figure4}(a)). The encapsulated sample is from sample A (bilayer in Fig.\ref{Figure2}(a)) and the non-encapsulated sample is from sample B (bilayer in Fig. \ref{Figure3}(a)). The computed spectra for the above sample structures are also plotted in Fig. \ref{Figure4}(b) to show that our model well accounts for the excitonic shift in different screening environments. A direct observation is that both the intralayer and interlayer exciton peaks become sharper upon encapsulation. Recently, there are studies\cite{cadiz_excitonic_2017,Ajayi_Approaching_2017} showing the hBN encapsulation provides a clean platform for high quality monolayer TMDC with narrow excitonic linewidth comparable with theoretical radiative broadening limit. We show that the encapsulation technique can also be applied to bilayer MoSe$_2$ for creating high quality samples. The intralayer exciton linewidth narrowing (from $46$~meV to $20$~meV for $A_{1s}$ state) can be explained by reduced inhomogeneity with hBN interfaces as well as a more symmetric dielectric screening provided by hBN encapsulation, which removes the energy difference between intralayer excitons dwells in the top and bottom layer. The observed interlayer exciton also exhibits linewidth narrowing from $35$~meV to $19$~meV, enabling us to identify the interlayer peak unambiguously. Note that the interlayer exciton in current work shows a narrower linewidth and a much larger oscillator strength than those observed in TMD heterojunctions, which typically have linewidth around 50-100 meV and oscillator strength 1/100 of the $A_{1s}$ state.\cite{michael_forg_cavity-control_nodate, miller_long-lived_2017,baranowski_probing_2017}

In summary, the strongly confined quasi-particles in vdWc layered semiconductors provide an interesting platform to form bound interlayer excitons. Through comparison between experimental and theoretical studies on linear optical spectra, we clearly identify the A-interlayer exciton in few-layer MoSe$_2$ systems. We first demonstrate that our DBE model reproduces the experimental monolayer MoSe$_2$ optical spectrum with a quantitative match and apply the same method to the few-layer MoSe$_2$ situation. The existence of interlayer excitons in layered single crystals substantiates the two-dimensional nature of electron behaviors, therefore our work is an essential step towards understanding the interlayer microscopic interactions in van der Waals materials.

The observed interlayer exciton has a large binding energy of $153$~meV, suggesting high thermal stability compared with conventional spatially indirect excitons, and a linewidth of $19$~meV, which is sharper than the interlayer exciton observed in TMD heterojunctions. Such strong Coulomb interaction in high quality vdWc materials suggest the possibility of using them as an interlayer exciton platform. Although the current interlayer exciton is in an indirect bandgap semiconductor, one can apply various methods to engineer the bandstructure of vdWc as well as the binding energy of excitons for creating long-lived interlayer excitons \cite{e.v._calman_indirect_nodate,zhao_continuously_2015,Lloyd_Band_2016}. Our current work will provide a cornerstone for future fundamental research in interlayer excitons of 2D materials.

{\bf Acknowledgment}
The experimental work in Michigan is supported by the United States Army Research Office MURI award W911NF-17-1-0312. The Marburg part of the work is a project of the Collaborative Research Center SFB 1083 funded by the Deutsche Forschungsgemeinschaft.

\bibliography{Refs}

\begin{thebibliography}{35}%
\makeatletter
\providecommand \@ifxundefined [1]{%
 \@ifx{#1\undefined}
}%
\providecommand \@ifnum [1]{%
 \ifnum #1\expandafter \@firstoftwo
 \else \expandafter \@secondoftwo
 \fi
}%
\providecommand \@ifx [1]{%
 \ifx #1\expandafter \@firstoftwo
 \else \expandafter \@secondoftwo
 \fi
}%
\providecommand \natexlab [1]{#1}%
\providecommand \enquote  [1]{``#1''}%
\providecommand \bibnamefont  [1]{#1}%
\providecommand \bibfnamefont [1]{#1}%
\providecommand \citenamefont [1]{#1}%
\providecommand \href@noop [0]{\@secondoftwo}%
\providecommand \href [0]{\begingroup \@sanitize@url \@href}%
\providecommand \@href[1]{\@@startlink{#1}\@@href}%
\providecommand \@@href[1]{\endgroup#1\@@endlink}%
\providecommand \@sanitize@url [0]{\catcode `\\12\catcode `\$12\catcode
  `\&12\catcode `\#12\catcode `\^12\catcode `\_12\catcode `\%12\relax}%
\providecommand \@@startlink[1]{}%
\providecommand \@@endlink[0]{}%
\providecommand \url  [0]{\begingroup\@sanitize@url \@url }%
\providecommand \@url [1]{\endgroup\@href {#1}{\urlprefix }}%
\providecommand \urlprefix  [0]{URL }%
\providecommand \Eprint [0]{\href }%
\providecommand \doibase [0]{http://dx.doi.org/}%
\providecommand \selectlanguage [0]{\@gobble}%
\providecommand \bibinfo  [0]{\@secondoftwo}%
\providecommand \bibfield  [0]{\@secondoftwo}%
\providecommand \translation [1]{[#1]}%
\providecommand \BibitemOpen [0]{}%
\providecommand \bibitemStop [0]{}%
\providecommand \bibitemNoStop [0]{.\EOS\space}%
\providecommand \EOS [0]{\spacefactor3000\relax}%
\providecommand \BibitemShut  [1]{\csname bibitem#1\endcsname}%
\let\auto@bib@innerbib\@empty
\bibitem [{\citenamefont {Frenkel}(1931)}]{frenkel_transformation_1931}%
  \BibitemOpen
  \bibfield  {author} {\bibinfo {author} {\bibfnamefont {J.}~\bibnamefont
  {Frenkel}},\ }\href {\doibase 10.1103/PhysRev.37.17} {\bibfield  {journal}
  {\bibinfo  {journal} {Physical Review}\ }\textbf {\bibinfo {volume} {37}},\
  \bibinfo {pages} {17} (\bibinfo {year} {1931})}\BibitemShut {NoStop}%
\bibitem [{\citenamefont {Wannier}(1937)}]{wannier_structure_1937}%
  \BibitemOpen
  \bibfield  {author} {\bibinfo {author} {\bibfnamefont {G.~H.}\ \bibnamefont
  {Wannier}},\ }\href {\doibase 10.1103/PhysRev.52.191} {\bibfield  {journal}
  {\bibinfo  {journal} {Physical Review}\ }\textbf {\bibinfo {volume} {52}},\
  \bibinfo {pages} {191} (\bibinfo {year} {1937})}\BibitemShut {NoStop}%
\bibitem [{\citenamefont {Colocci}\ \emph {et~al.}(1990)\citenamefont
  {Colocci}, \citenamefont {Gurioli}, \citenamefont {Vinattieri}, \citenamefont
  {Fermi}, \citenamefont {Deparis}, \citenamefont {Massies},\ and\
  \citenamefont {Neu}}]{Colocci_Temperature_1990}%
  \BibitemOpen
  \bibfield  {author} {\bibinfo {author} {\bibfnamefont {M.}~\bibnamefont
  {Colocci}}, \bibinfo {author} {\bibfnamefont {M.}~\bibnamefont {Gurioli}},
  \bibinfo {author} {\bibfnamefont {A.}~\bibnamefont {Vinattieri}}, \bibinfo
  {author} {\bibfnamefont {F.}~\bibnamefont {Fermi}}, \bibinfo {author}
  {\bibfnamefont {C.}~\bibnamefont {Deparis}}, \bibinfo {author} {\bibfnamefont
  {J.}~\bibnamefont {Massies}}, \ and\ \bibinfo {author} {\bibfnamefont
  {G.}~\bibnamefont {Neu}},\ }\href
  {http://stacks.iop.org/0295-5075/12/i=5/a=007} {\bibfield  {journal}
  {\bibinfo  {journal} {EPL (Europhysics Letters)}\ }\textbf {\bibinfo {volume}
  {12}},\ \bibinfo {pages} {417} (\bibinfo {year} {1990})}\BibitemShut
  {NoStop}%
\bibitem [{\citenamefont {Butov}\ \emph {et~al.}(1994)\citenamefont {Butov},
  \citenamefont {Zrenner}, \citenamefont {Abstreiter}, \citenamefont {B\"ohm},\
  and\ \citenamefont {Weimann}}]{Butov_Condensation_1994}%
  \BibitemOpen
  \bibfield  {author} {\bibinfo {author} {\bibfnamefont {L.~V.}\ \bibnamefont
  {Butov}}, \bibinfo {author} {\bibfnamefont {A.}~\bibnamefont {Zrenner}},
  \bibinfo {author} {\bibfnamefont {G.}~\bibnamefont {Abstreiter}}, \bibinfo
  {author} {\bibfnamefont {G.}~\bibnamefont {B\"ohm}}, \ and\ \bibinfo {author}
  {\bibfnamefont {G.}~\bibnamefont {Weimann}},\ }\href {\doibase
  10.1103/PhysRevLett.73.304} {\bibfield  {journal} {\bibinfo  {journal} {Phys.
  Rev. Lett.}\ }\textbf {\bibinfo {volume} {73}},\ \bibinfo {pages} {304}
  (\bibinfo {year} {1994})}\BibitemShut {NoStop}%
\bibitem [{\citenamefont {Geim}\ and\ \citenamefont
  {Grigorieva}(2013)}]{geim_van_2013}%
  \BibitemOpen
  \bibfield  {author} {\bibinfo {author} {\bibfnamefont {A.~K.}\ \bibnamefont
  {Geim}}\ and\ \bibinfo {author} {\bibfnamefont {I.~V.}\ \bibnamefont
  {Grigorieva}},\ }\href {http://dx.doi.org/10.1038/nature12385} {\bibfield
  {journal} {\bibinfo  {journal} {Nature}\ }\textbf {\bibinfo {volume} {499}},\
  \bibinfo {pages} {419} (\bibinfo {year} {2013})}\BibitemShut {NoStop}%
\bibitem [{\citenamefont {Jariwala}\ \emph {et~al.}(2016)\citenamefont
  {Jariwala}, \citenamefont {Marks},\ and\ \citenamefont
  {Hersam}}]{jariwala_mixed-dimensional_2016}%
  \BibitemOpen
  \bibfield  {author} {\bibinfo {author} {\bibfnamefont {D.}~\bibnamefont
  {Jariwala}}, \bibinfo {author} {\bibfnamefont {T.~J.}\ \bibnamefont {Marks}},
  \ and\ \bibinfo {author} {\bibfnamefont {M.~C.}\ \bibnamefont {Hersam}},\
  }\href {http://dx.doi.org/10.1038/nmat4703} {\bibfield  {journal} {\bibinfo
  {journal} {Nature Materials}\ }\textbf {\bibinfo {volume} {16}},\ \bibinfo
  {pages} {170} (\bibinfo {year} {2016})}\BibitemShut {NoStop}%
\bibitem [{\citenamefont {Rivera}\ \emph {et~al.}(2015)\citenamefont {Rivera},
  \citenamefont {Schaibley}, \citenamefont {Jones}, \citenamefont {Ross},
  \citenamefont {Wu}, \citenamefont {Aivazian}, \citenamefont {Klement},
  \citenamefont {Seyler}, \citenamefont {Clark}, \citenamefont {Ghimire},
  \citenamefont {Yan}, \citenamefont {Mandrus}, \citenamefont {Yao},\ and\
  \citenamefont {Xu}}]{rivera_observation_2015}%
  \BibitemOpen
  \bibfield  {author} {\bibinfo {author} {\bibfnamefont {P.}~\bibnamefont
  {Rivera}}, \bibinfo {author} {\bibfnamefont {J.~R.}\ \bibnamefont
  {Schaibley}}, \bibinfo {author} {\bibfnamefont {A.~M.}\ \bibnamefont
  {Jones}}, \bibinfo {author} {\bibfnamefont {J.~S.}\ \bibnamefont {Ross}},
  \bibinfo {author} {\bibfnamefont {S.}~\bibnamefont {Wu}}, \bibinfo {author}
  {\bibfnamefont {G.}~\bibnamefont {Aivazian}}, \bibinfo {author}
  {\bibfnamefont {P.}~\bibnamefont {Klement}}, \bibinfo {author} {\bibfnamefont
  {K.}~\bibnamefont {Seyler}}, \bibinfo {author} {\bibfnamefont
  {G.}~\bibnamefont {Clark}}, \bibinfo {author} {\bibfnamefont {N.~J.}\
  \bibnamefont {Ghimire}}, \bibinfo {author} {\bibfnamefont {J.}~\bibnamefont
  {Yan}}, \bibinfo {author} {\bibfnamefont {D.~G.}\ \bibnamefont {Mandrus}},
  \bibinfo {author} {\bibfnamefont {W.}~\bibnamefont {Yao}}, \ and\ \bibinfo
  {author} {\bibfnamefont {X.}~\bibnamefont {Xu}},\ }\href
  {http://dx.doi.org/10.1038/ncomms7242} {\bibfield  {journal} {\bibinfo
  {journal} {Nature Communications}\ }\textbf {\bibinfo {volume} {6}},\
  \bibinfo {pages} {6242} (\bibinfo {year} {2015})}\BibitemShut {NoStop}%
\bibitem [{\citenamefont {{Philipp Nagler}}\ \emph {et~al.}(2017)\citenamefont
  {{Philipp Nagler}}, \citenamefont {{Gerd Plechinger}}, \citenamefont
  {{Mariana V Ballottin}}, \citenamefont {{Anatolie Mitioglu}}, \citenamefont
  {{Sebastian Meier}}, \citenamefont {{Nicola Paradiso}}, \citenamefont
  {{Christoph Strunk}}, \citenamefont {{Alexey Chernikov}}, \citenamefont
  {{Peter C M Christianen}}, \citenamefont {{Christian Schüller}},\ and\
  \citenamefont {{Tobias Korn}}}]{philipp_nagler_interlayer_2017}%
  \BibitemOpen
  \bibfield  {author} {\bibinfo {author} {\bibnamefont {{Philipp Nagler}}},
  \bibinfo {author} {\bibnamefont {{Gerd Plechinger}}}, \bibinfo {author}
  {\bibnamefont {{Mariana V Ballottin}}}, \bibinfo {author} {\bibnamefont
  {{Anatolie Mitioglu}}}, \bibinfo {author} {\bibnamefont {{Sebastian Meier}}},
  \bibinfo {author} {\bibnamefont {{Nicola Paradiso}}}, \bibinfo {author}
  {\bibnamefont {{Christoph Strunk}}}, \bibinfo {author} {\bibnamefont {{Alexey
  Chernikov}}}, \bibinfo {author} {\bibnamefont {{Peter C M Christianen}}},
  \bibinfo {author} {\bibnamefont {{Christian Schüller}}}, \ and\ \bibinfo
  {author} {\bibnamefont {{Tobias Korn}}},\ }\href
  {http://stacks.iop.org/2053-1583/4/i=2/a=025112} {\bibfield  {journal}
  {\bibinfo  {journal} {2D Materials}\ }\textbf {\bibinfo {volume} {4}},\
  \bibinfo {pages} {025112} (\bibinfo {year} {2017})}\BibitemShut {NoStop}%
\bibitem [{\citenamefont {Zhu}\ \emph {et~al.}(2015)\citenamefont {Zhu},
  \citenamefont {Monahan}, \citenamefont {Gong}, \citenamefont {Zhu},
  \citenamefont {Williams},\ and\ \citenamefont {Nelson}}]{zhu_charge_2015}%
  \BibitemOpen
  \bibfield  {author} {\bibinfo {author} {\bibfnamefont {X.}~\bibnamefont
  {Zhu}}, \bibinfo {author} {\bibfnamefont {N.~R.}\ \bibnamefont {Monahan}},
  \bibinfo {author} {\bibfnamefont {Z.}~\bibnamefont {Gong}}, \bibinfo {author}
  {\bibfnamefont {H.}~\bibnamefont {Zhu}}, \bibinfo {author} {\bibfnamefont
  {K.~W.}\ \bibnamefont {Williams}}, \ and\ \bibinfo {author} {\bibfnamefont
  {C.~A.}\ \bibnamefont {Nelson}},\ }\href {\doibase 10.1021/jacs.5b03141}
  {\bibfield  {journal} {\bibinfo  {journal} {Journal of the American Chemical
  Society}\ }\textbf {\bibinfo {volume} {137}},\ \bibinfo {pages} {8313}
  (\bibinfo {year} {2015})}\BibitemShut {NoStop}%
\bibitem [{\citenamefont {Schaibley}\ \emph {et~al.}(2016)\citenamefont
  {Schaibley}, \citenamefont {Rivera}, \citenamefont {Yu}, \citenamefont
  {Seyler}, \citenamefont {Yan}, \citenamefont {Mandrus}, \citenamefont
  {Taniguchi}, \citenamefont {Watanabe}, \citenamefont {Yao},\ and\
  \citenamefont {Xu}}]{schaibley_directional_2016}%
  \BibitemOpen
  \bibfield  {author} {\bibinfo {author} {\bibfnamefont {J.~R.}\ \bibnamefont
  {Schaibley}}, \bibinfo {author} {\bibfnamefont {P.}~\bibnamefont {Rivera}},
  \bibinfo {author} {\bibfnamefont {H.}~\bibnamefont {Yu}}, \bibinfo {author}
  {\bibfnamefont {K.~L.}\ \bibnamefont {Seyler}}, \bibinfo {author}
  {\bibfnamefont {J.}~\bibnamefont {Yan}}, \bibinfo {author} {\bibfnamefont
  {D.~G.}\ \bibnamefont {Mandrus}}, \bibinfo {author} {\bibfnamefont
  {T.}~\bibnamefont {Taniguchi}}, \bibinfo {author} {\bibfnamefont
  {K.}~\bibnamefont {Watanabe}}, \bibinfo {author} {\bibfnamefont
  {W.}~\bibnamefont {Yao}}, \ and\ \bibinfo {author} {\bibfnamefont
  {X.}~\bibnamefont {Xu}},\ }\href {http://dx.doi.org/10.1038/ncomms13747}
  {\bibfield  {journal} {\bibinfo  {journal} {Nature Communications}\ }\textbf
  {\bibinfo {volume} {7}},\ \bibinfo {pages} {13747} (\bibinfo {year}
  {2016})}\BibitemShut {NoStop}%
\bibitem [{\citenamefont {Ross}\ \emph {et~al.}(2017)\citenamefont {Ross},
  \citenamefont {Rivera}, \citenamefont {Schaibley}, \citenamefont {Lee-Wong},
  \citenamefont {Yu}, \citenamefont {Taniguchi}, \citenamefont {Watanabe},
  \citenamefont {Yan}, \citenamefont {Mandrus}, \citenamefont {Cobden},
  \citenamefont {Yao},\ and\ \citenamefont {Xu}}]{ross_interlayer_2017}%
  \BibitemOpen
  \bibfield  {author} {\bibinfo {author} {\bibfnamefont {J.~S.}\ \bibnamefont
  {Ross}}, \bibinfo {author} {\bibfnamefont {P.}~\bibnamefont {Rivera}},
  \bibinfo {author} {\bibfnamefont {J.}~\bibnamefont {Schaibley}}, \bibinfo
  {author} {\bibfnamefont {E.}~\bibnamefont {Lee-Wong}}, \bibinfo {author}
  {\bibfnamefont {H.}~\bibnamefont {Yu}}, \bibinfo {author} {\bibfnamefont
  {T.}~\bibnamefont {Taniguchi}}, \bibinfo {author} {\bibfnamefont
  {K.}~\bibnamefont {Watanabe}}, \bibinfo {author} {\bibfnamefont
  {J.}~\bibnamefont {Yan}}, \bibinfo {author} {\bibfnamefont {D.}~\bibnamefont
  {Mandrus}}, \bibinfo {author} {\bibfnamefont {D.}~\bibnamefont {Cobden}},
  \bibinfo {author} {\bibfnamefont {W.}~\bibnamefont {Yao}}, \ and\ \bibinfo
  {author} {\bibfnamefont {X.}~\bibnamefont {Xu}},\ }\href {\doibase
  10.1021/acs.nanolett.6b03398} {\bibfield  {journal} {\bibinfo  {journal}
  {Nano Letters}\ }\textbf {\bibinfo {volume} {17}},\ \bibinfo {pages} {638}
  (\bibinfo {year} {2017})}\BibitemShut {NoStop}%
\bibitem [{\citenamefont {Meckbach}\ \emph
  {et~al.}(2017{\natexlab{a}})\citenamefont {Meckbach}, \citenamefont
  {Stroucken},\ and\ \citenamefont {Koch}}]{Meckbach2017a}%
  \BibitemOpen
  \bibfield  {author} {\bibinfo {author} {\bibfnamefont {L.}~\bibnamefont
  {Meckbach}}, \bibinfo {author} {\bibfnamefont {T.}~\bibnamefont {Stroucken}},
  \ and\ \bibinfo {author} {\bibfnamefont {S.~W.}\ \bibnamefont {Koch}},\
  }\href {https://arxiv.org/abs/1709.09056} {\bibfield  {journal} {\bibinfo
  {journal} {arXiv:1709.09056}\ } (\bibinfo {year}
  {2017}{\natexlab{a}})}\BibitemShut {NoStop}%
\bibitem [{\citenamefont {Meckbach}\ \emph
  {et~al.}(2017{\natexlab{b}})\citenamefont {Meckbach}, \citenamefont
  {Stroucken},\ and\ \citenamefont {Koch}}]{Meckbach2017b}%
  \BibitemOpen
  \bibfield  {author} {\bibinfo {author} {\bibfnamefont {L.}~\bibnamefont
  {Meckbach}}, \bibinfo {author} {\bibfnamefont {T.}~\bibnamefont {Stroucken}},
  \ and\ \bibinfo {author} {\bibfnamefont {S.~W.}\ \bibnamefont {Koch}},\
  }\href {https://arxiv.org/abs/1709.09056} {\bibfield  {journal} {\bibinfo
  {journal} {Submitted to Apllied Physics Letters}\ } (\bibinfo {year}
  {2017}{\natexlab{b}})}\BibitemShut {NoStop}%
\bibitem [{\citenamefont {Fogler}\ \emph {et~al.}(2014)\citenamefont {Fogler},
  \citenamefont {Butov},\ and\ \citenamefont
  {Novoselov}}]{fogler_high-temperature_2014}%
  \BibitemOpen
  \bibfield  {author} {\bibinfo {author} {\bibfnamefont {M.~M.}\ \bibnamefont
  {Fogler}}, \bibinfo {author} {\bibfnamefont {L.~V.}\ \bibnamefont {Butov}}, \
  and\ \bibinfo {author} {\bibfnamefont {K.~S.}\ \bibnamefont {Novoselov}},\
  }\href {http://dx.doi.org/10.1038/ncomms5555} {\bibfield  {journal} {\bibinfo
   {journal} {Nature Communications}\ }\textbf {\bibinfo {volume} {5}},\
  \bibinfo {pages} {4555} (\bibinfo {year} {2014})}\BibitemShut {NoStop}%
\bibitem [{\citenamefont {{E.V. Calman}}()}]{e.v._calman_indirect_nodate}%
  \BibitemOpen
  \bibfield  {author} {\bibinfo {author} {\bibnamefont {{E.V. Calman}}},\
  }\href@noop {} {\bibinfo  {journal} {arXiv:1709.07043v1}\ }\BibitemShut
  {NoStop}%
\bibitem [{\citenamefont {{Mathieu Alloing}}\ \emph {et~al.}(2014)\citenamefont
  {{Mathieu Alloing}}, \citenamefont {{Mussie Beian}}, \citenamefont {{Maciej
  Lewenstein}}, \citenamefont {{David Fuster}}, \citenamefont {{Yolanda
  González}}, \citenamefont {{Luisa González}}, \citenamefont {{Roland
  Combescot}}, \citenamefont {{Monique Combescot}},\ and\ \citenamefont
  {{François Dubin}}}]{mathieu_alloing_evidence_2014}%
  \BibitemOpen
\bibfield  {journal} {  }\bibfield  {author} {\bibinfo {author} {\bibnamefont
  {{Mathieu Alloing}}}, \bibinfo {author} {\bibnamefont {{Mussie Beian}}},
  \bibinfo {author} {\bibnamefont {{Maciej Lewenstein}}}, \bibinfo {author}
  {\bibnamefont {{David Fuster}}}, \bibinfo {author} {\bibnamefont {{Yolanda
  González}}}, \bibinfo {author} {\bibnamefont {{Luisa González}}}, \bibinfo
  {author} {\bibnamefont {{Roland Combescot}}}, \bibinfo {author} {\bibnamefont
  {{Monique Combescot}}}, \ and\ \bibinfo {author} {\bibnamefont {{François
  Dubin}}},\ }\href {http://stacks.iop.org/0295-5075/107/i=1/a=10012}
  {\bibfield  {journal} {\bibinfo  {journal} {EPL (Europhysics Letters)}\
  }\textbf {\bibinfo {volume} {107}},\ \bibinfo {pages} {10012} (\bibinfo
  {year} {2014})}\BibitemShut {NoStop}%
\bibitem [{\citenamefont {{Monique Combescot}}\ \emph
  {et~al.}(2017)\citenamefont {{Monique Combescot}}, \citenamefont {{Roland
  Combescot}},\ and\ \citenamefont {{François
  Dubin}}}]{monique_combescot_boseeinstein_2017}%
  \BibitemOpen
  \bibfield  {author} {\bibinfo {author} {\bibnamefont {{Monique Combescot}}},
  \bibinfo {author} {\bibnamefont {{Roland Combescot}}}, \ and\ \bibinfo
  {author} {\bibnamefont {{François Dubin}}},\ }\href
  {http://stacks.iop.org/0034-4885/80/i=6/a=066501} {\bibfield  {journal}
  {\bibinfo  {journal} {Reports on Progress in Physics}\ }\textbf {\bibinfo
  {volume} {80}},\ \bibinfo {pages} {066501} (\bibinfo {year}
  {2017})}\BibitemShut {NoStop}%
\bibitem [{\citenamefont {Su}\ and\ \citenamefont
  {MacDonald}(2017)}]{su_spatially_2017}%
  \BibitemOpen
  \bibfield  {author} {\bibinfo {author} {\bibfnamefont {J.-J.}\ \bibnamefont
  {Su}}\ and\ \bibinfo {author} {\bibfnamefont {A.~H.}\ \bibnamefont
  {MacDonald}},\ }\href {\doibase 10.1103/PhysRevB.95.045416} {\bibfield
  {journal} {\bibinfo  {journal} {Phys. Rev. B}\ }\textbf {\bibinfo {volume}
  {95}},\ \bibinfo {pages} {045416} (\bibinfo {year} {2017})}\BibitemShut
  {NoStop}%
\bibitem [{\citenamefont {Berman}\ and\ \citenamefont
  {Kezerashvili}(2017)}]{berman_superfluidity_2017}%
  \BibitemOpen
  \bibfield  {author} {\bibinfo {author} {\bibfnamefont {O.~L.}\ \bibnamefont
  {Berman}}\ and\ \bibinfo {author} {\bibfnamefont {R.~Y.}\ \bibnamefont
  {Kezerashvili}},\ }\href {\doibase 10.1103/PhysRevB.96.094502} {\bibfield
  {journal} {\bibinfo  {journal} {Phys. Rev. B}\ }\textbf {\bibinfo {volume}
  {96}},\ \bibinfo {pages} {094502} (\bibinfo {year} {2017})}\BibitemShut
  {NoStop}%
\bibitem [{\citenamefont {Deng}\ \emph {et~al.}(2010)\citenamefont {Deng},
  \citenamefont {Haug},\ and\ \citenamefont
  {Yamamoto}}]{deng_exciton-polariton_2010}%
  \BibitemOpen
  \bibfield  {author} {\bibinfo {author} {\bibfnamefont {H.}~\bibnamefont
  {Deng}}, \bibinfo {author} {\bibfnamefont {H.}~\bibnamefont {Haug}}, \ and\
  \bibinfo {author} {\bibfnamefont {Y.}~\bibnamefont {Yamamoto}},\ }\href
  {\doibase 10.1103/RevModPhys.82.1489} {\bibfield  {journal} {\bibinfo
  {journal} {Reviews of Modern Physics}\ }\textbf {\bibinfo {volume} {82}},\
  \bibinfo {pages} {1489} (\bibinfo {year} {2010})}\BibitemShut {NoStop}%
\bibitem [{\citenamefont {Carusotto}\ and\ \citenamefont
  {Ciuti}(2013)}]{carusotto_quantum_2013}%
  \BibitemOpen
  \bibfield  {author} {\bibinfo {author} {\bibfnamefont {I.}~\bibnamefont
  {Carusotto}}\ and\ \bibinfo {author} {\bibfnamefont {C.}~\bibnamefont
  {Ciuti}},\ }\href {\doibase 10.1103/RevModPhys.85.299} {\bibfield  {journal}
  {\bibinfo  {journal} {Reviews of Modern Physics}\ }\textbf {\bibinfo {volume}
  {85}},\ \bibinfo {pages} {299} (\bibinfo {year} {2013})}\BibitemShut
  {NoStop}%
\bibitem [{\citenamefont {Ye}\ \emph {et~al.}(2015)\citenamefont {Ye},
  \citenamefont {Winslow}, \citenamefont {Zhang}, \citenamefont {Pandey},\ and\
  \citenamefont {Yap}}]{ye2015}%
  \BibitemOpen
  \bibfield  {author} {\bibinfo {author} {\bibfnamefont {M.}~\bibnamefont
  {Ye}}, \bibinfo {author} {\bibfnamefont {D.}~\bibnamefont {Winslow}},
  \bibinfo {author} {\bibfnamefont {D.}~\bibnamefont {Zhang}}, \bibinfo
  {author} {\bibfnamefont {R.}~\bibnamefont {Pandey}}, \ and\ \bibinfo {author}
  {\bibfnamefont {Y.~K.}\ \bibnamefont {Yap}},\ }\href {\doibase
  10.3390/photonics2010288} {\bibfield  {journal} {\bibinfo  {journal}
  {Photonics}\ }\textbf {\bibinfo {volume} {2}},\ \bibinfo {pages} {288 }
  (\bibinfo {year} {2015})}\BibitemShut {NoStop}%
\bibitem [{\citenamefont {Xu}\ \emph {et~al.}(2014)\citenamefont {Xu},
  \citenamefont {Yao}, \citenamefont {Xiao},\ and\ \citenamefont
  {Heinz}}]{xu_spin_2014}%
  \BibitemOpen
  \bibfield  {author} {\bibinfo {author} {\bibfnamefont {X.}~\bibnamefont
  {Xu}}, \bibinfo {author} {\bibfnamefont {W.}~\bibnamefont {Yao}}, \bibinfo
  {author} {\bibfnamefont {D.}~\bibnamefont {Xiao}}, \ and\ \bibinfo {author}
  {\bibfnamefont {T.~F.}\ \bibnamefont {Heinz}},\ }\href
  {http://dx.doi.org/10.1038/nphys2942} {\bibfield  {journal} {\bibinfo
  {journal} {Nature Physics}\ }\textbf {\bibinfo {volume} {10}},\ \bibinfo
  {pages} {343} (\bibinfo {year} {2014})}\BibitemShut {NoStop}%
\bibitem [{\citenamefont {Xiao}\ \emph {et~al.}(2012)\citenamefont {Xiao},
  \citenamefont {Liu}, \citenamefont {Feng}, \citenamefont {Xu},\ and\
  \citenamefont {Yao}}]{xiao2012}%
  \BibitemOpen
  \bibfield  {author} {\bibinfo {author} {\bibfnamefont {D.}~\bibnamefont
  {Xiao}}, \bibinfo {author} {\bibfnamefont {G.-B.}\ \bibnamefont {Liu}},
  \bibinfo {author} {\bibfnamefont {W.}~\bibnamefont {Feng}}, \bibinfo {author}
  {\bibfnamefont {X.}~\bibnamefont {Xu}}, \ and\ \bibinfo {author}
  {\bibfnamefont {W.}~\bibnamefont {Yao}},\ }\href {\doibase
  10.1103/PhysRevLett.108.196802} {\bibfield  {journal} {\bibinfo  {journal}
  {Phys. Rev. Lett.}\ }\textbf {\bibinfo {volume} {108}},\ \bibinfo {pages}
  {196802} (\bibinfo {year} {2012})}\BibitemShut {NoStop}%
\bibitem [{not()}]{note}%
  \BibitemOpen
  \href@noop {} {}\bibinfo {note} {In the massive Dirac-Fermion for MoSe$_2$
  layers, we use the parameters the energy gap $\Delta=1.47$ eV, the effetive
  hopping matrix element $t=0.94 eV$, the lattice constant $a=3.313 \,\AA$ and
  the spin splitting of valence and conduction band $2\lambda_\nu=0.18$ eV and
  $2\lambda_c=0.02$ eV, as given in Ref.\cite{xiao2012,kosmider2013}. The
  in-plane and out-of-plane background dielectric constants are 3.36 and 5.16,
  respectively. Furthermore, we use a natural layer-to-layer distance of
  $D=6.5\,\AA$ and an effective thickness parameter characterizing the finite
  extension of the Mo-$d$-orbitals in out-of-plane direction of
  $5\,\AA$.}\BibitemShut {Stop}%
\bibitem [{\citenamefont {Riley}\ \emph {et~al.}(2014)\citenamefont {Riley},
  \citenamefont {Mazzola}, \citenamefont {Dendzik}, \citenamefont {Michiardi},
  \citenamefont {Takayama}, \citenamefont {Bawden}, \citenamefont {Graner�d},
  \citenamefont {Leandersson}, \citenamefont {Balasubramanian}, \citenamefont
  {Hoesch}, \citenamefont {Kim}, \citenamefont {Takagi}, \citenamefont
  {Meevasana}, \citenamefont {Hofmann}, \citenamefont {Bahramy}, \citenamefont
  {Wells},\ and\ \citenamefont {King}}]{Riley2014}%
  \BibitemOpen
  \bibfield  {author} {\bibinfo {author} {\bibfnamefont {J.~M.}\ \bibnamefont
  {Riley}}, \bibinfo {author} {\bibfnamefont {F.}~\bibnamefont {Mazzola}},
  \bibinfo {author} {\bibfnamefont {M.}~\bibnamefont {Dendzik}}, \bibinfo
  {author} {\bibfnamefont {M.}~\bibnamefont {Michiardi}}, \bibinfo {author}
  {\bibfnamefont {T.}~\bibnamefont {Takayama}}, \bibinfo {author}
  {\bibfnamefont {L.}~\bibnamefont {Bawden}}, \bibinfo {author} {\bibfnamefont
  {C.}~\bibnamefont {Graner�d}}, \bibinfo {author} {\bibfnamefont
  {M.}~\bibnamefont {Leandersson}}, \bibinfo {author} {\bibfnamefont
  {T.}~\bibnamefont {Balasubramanian}}, \bibinfo {author} {\bibfnamefont
  {M.}~\bibnamefont {Hoesch}}, \bibinfo {author} {\bibfnamefont {T.~K.}\
  \bibnamefont {Kim}}, \bibinfo {author} {\bibfnamefont {H.}~\bibnamefont
  {Takagi}}, \bibinfo {author} {\bibfnamefont {W.}~\bibnamefont {Meevasana}},
  \bibinfo {author} {\bibfnamefont {P.}~\bibnamefont {Hofmann}}, \bibinfo
  {author} {\bibfnamefont {M.~S.}\ \bibnamefont {Bahramy}}, \bibinfo {author}
  {\bibfnamefont {J.~W.}\ \bibnamefont {Wells}}, \ and\ \bibinfo {author}
  {\bibfnamefont {P.~D.~C.}\ \bibnamefont {King}},\ }\href@noop {} {\bibfield
  {journal} {\bibinfo  {journal} {Nature Physics}\ }\textbf {\bibinfo {volume}
  {10}},\ \bibinfo {pages} {835} (\bibinfo {year} {2014})}\BibitemShut
  {NoStop}%
\bibitem [{\citenamefont {Haug}\ and\ \citenamefont
  {Koch}(2009)}]{haugkoch2009}%
  \BibitemOpen
  \bibfield  {author} {\bibinfo {author} {\bibfnamefont {H.}~\bibnamefont
  {Haug}}\ and\ \bibinfo {author} {\bibfnamefont {S.~W.}\ \bibnamefont
  {Koch}},\ }\href {http://books.google.de/books?id=qHx2YWx1LOcC} {\emph
  {\bibinfo {title} {{Quantum Theory of the Optical and Electronic Properties
  of Semiconductors}}}},\ \bibinfo {edition} {5th}\ ed.\ (\bibinfo  {publisher}
  {World Scientific Publishing},\ \bibinfo {address} {Singapur},\ \bibinfo
  {year} {2009})\BibitemShut {NoStop}%
\bibitem [{\citenamefont {Chernikov}\ \emph {et~al.}(2014)\citenamefont
  {Chernikov}, \citenamefont {Berkelbach}, \citenamefont {Hill}, \citenamefont
  {Rigosi}, \citenamefont {Li},\ and\ \citenamefont
  {Aslan}}]{chernikov_exciton_2014}%
  \BibitemOpen
  \bibfield  {author} {\bibinfo {author} {\bibfnamefont {A.}~\bibnamefont
  {Chernikov}}, \bibinfo {author} {\bibfnamefont {T.~C.}\ \bibnamefont
  {Berkelbach}}, \bibinfo {author} {\bibfnamefont {H.~M.}\ \bibnamefont
  {Hill}}, \bibinfo {author} {\bibfnamefont {A.}~\bibnamefont {Rigosi}},
  \bibinfo {author} {\bibfnamefont {Y.}~\bibnamefont {Li}}, \ and\ \bibinfo
  {author} {\bibfnamefont {O.~B.}\ \bibnamefont {Aslan}},\ }\href {\doibase
  10.1103/PhysRevLett.113.076802} {\bibfield  {journal} {\bibinfo  {journal}
  {Phys. Rev. Lett.}\ }\textbf {\bibinfo {volume} {113}},\ \bibinfo {pages}
  {076802} (\bibinfo {year} {2014})}\BibitemShut {NoStop}%
\bibitem [{\citenamefont {Cadiz}\ \emph {et~al.}(2017)\citenamefont {Cadiz},
  \citenamefont {Courtade}, \citenamefont {Robert}, \citenamefont {Wang},
  \citenamefont {Shen}, \citenamefont {Cai}, \citenamefont {Taniguchi},
  \citenamefont {Watanabe}, \citenamefont {Carrere}, \citenamefont {Lagarde},
  \citenamefont {Manca}, \citenamefont {Amand}, \citenamefont {Renucci},
  \citenamefont {Tongay}, \citenamefont {Marie},\ and\ \citenamefont
  {Urbaszek}}]{cadiz_excitonic_2017}%
  \BibitemOpen
  \bibfield  {author} {\bibinfo {author} {\bibfnamefont {F.}~\bibnamefont
  {Cadiz}}, \bibinfo {author} {\bibfnamefont {E.}~\bibnamefont {Courtade}},
  \bibinfo {author} {\bibfnamefont {C.}~\bibnamefont {Robert}}, \bibinfo
  {author} {\bibfnamefont {G.}~\bibnamefont {Wang}}, \bibinfo {author}
  {\bibfnamefont {Y.}~\bibnamefont {Shen}}, \bibinfo {author} {\bibfnamefont
  {H.}~\bibnamefont {Cai}}, \bibinfo {author} {\bibfnamefont {T.}~\bibnamefont
  {Taniguchi}}, \bibinfo {author} {\bibfnamefont {K.}~\bibnamefont {Watanabe}},
  \bibinfo {author} {\bibfnamefont {H.}~\bibnamefont {Carrere}}, \bibinfo
  {author} {\bibfnamefont {D.}~\bibnamefont {Lagarde}}, \bibinfo {author}
  {\bibfnamefont {M.}~\bibnamefont {Manca}}, \bibinfo {author} {\bibfnamefont
  {T.}~\bibnamefont {Amand}}, \bibinfo {author} {\bibfnamefont
  {P.}~\bibnamefont {Renucci}}, \bibinfo {author} {\bibfnamefont
  {S.}~\bibnamefont {Tongay}}, \bibinfo {author} {\bibfnamefont
  {X.}~\bibnamefont {Marie}}, \ and\ \bibinfo {author} {\bibfnamefont
  {B.}~\bibnamefont {Urbaszek}},\ }\href {\doibase 10.1103/PhysRevX.7.021026}
  {\bibfield  {journal} {\bibinfo  {journal} {Phys. Rev. X}\ }\textbf {\bibinfo
  {volume} {7}},\ \bibinfo {pages} {021026} (\bibinfo {year}
  {2017})}\BibitemShut {NoStop}%
\bibitem [{\citenamefont {Ajayi}\ \emph {et~al.}(2017)\citenamefont {Ajayi},
  \citenamefont {Ardelean}, \citenamefont {Shepard}, \citenamefont {Wang},
  \citenamefont {Antony}, \citenamefont {Taniguchi}, \citenamefont {Watanabe},
  \citenamefont {Heinz}, \citenamefont {Strauf}, \citenamefont {Zhu},\ and\
  \citenamefont {Hone}}]{Ajayi_Approaching_2017}%
  \BibitemOpen
  \bibfield  {author} {\bibinfo {author} {\bibfnamefont {O.~A.}\ \bibnamefont
  {Ajayi}}, \bibinfo {author} {\bibfnamefont {J.~V.}\ \bibnamefont {Ardelean}},
  \bibinfo {author} {\bibfnamefont {G.~D.}\ \bibnamefont {Shepard}}, \bibinfo
  {author} {\bibfnamefont {J.}~\bibnamefont {Wang}}, \bibinfo {author}
  {\bibfnamefont {A.}~\bibnamefont {Antony}}, \bibinfo {author} {\bibfnamefont
  {T.}~\bibnamefont {Taniguchi}}, \bibinfo {author} {\bibfnamefont
  {K.}~\bibnamefont {Watanabe}}, \bibinfo {author} {\bibfnamefont {T.~F.}\
  \bibnamefont {Heinz}}, \bibinfo {author} {\bibfnamefont {S.}~\bibnamefont
  {Strauf}}, \bibinfo {author} {\bibfnamefont {X.-Y.}\ \bibnamefont {Zhu}}, \
  and\ \bibinfo {author} {\bibfnamefont {J.~C.}\ \bibnamefont {Hone}},\ }\href
  {http://stacks.iop.org/2053-1583/4/i=3/a=031011} {\bibfield  {journal}
  {\bibinfo  {journal} {2D Materials}\ }\textbf {\bibinfo {volume} {4}},\
  \bibinfo {pages} {031011} (\bibinfo {year} {2017})}\BibitemShut {NoStop}%
\bibitem [{\citenamefont {{Michael Förg}}\ \emph {et~al.}()\citenamefont
  {{Michael Förg}}, \citenamefont {{Léo Colombier}}, \citenamefont {{Robin K.
  Patel}}, \citenamefont {{Jessica Lindlau}}, \citenamefont {{Aditya D.
  Mohite}}, \citenamefont {{Hisato Yamaguchi}}, \citenamefont {{David
  Hunger}},\ and\ \citenamefont {{Alexander
  Högele}}}]{michael_forg_cavity-control_nodate}%
  \BibitemOpen
  \bibfield  {author} {\bibinfo {author} {\bibnamefont {{Michael Förg}}},
  \bibinfo {author} {\bibnamefont {{Léo Colombier}}}, \bibinfo {author}
  {\bibnamefont {{Robin K. Patel}}}, \bibinfo {author} {\bibnamefont {{Jessica
  Lindlau}}}, \bibinfo {author} {\bibnamefont {{Aditya D. Mohite}}}, \bibinfo
  {author} {\bibnamefont {{Hisato Yamaguchi}}}, \bibinfo {author} {\bibnamefont
  {{David Hunger}}}, \ and\ \bibinfo {author} {\bibnamefont {{Alexander
  Högele}}},\ }\href@noop {} {\bibinfo  {journal} {arXiv:1710.00990v2}\
  }\BibitemShut {NoStop}%
\bibitem [{\citenamefont {Miller}\ \emph {et~al.}(2017)\citenamefont {Miller},
  \citenamefont {Steinhoff}, \citenamefont {Pano}, \citenamefont {Klein},
  \citenamefont {Jahnke}, \citenamefont {Holleitner},\ and\ \citenamefont
  {Wurstbauer}}]{miller_long-lived_2017}%
  \BibitemOpen
\bibfield  {journal} {  }\bibfield  {author} {\bibinfo {author} {\bibfnamefont
  {B.}~\bibnamefont {Miller}}, \bibinfo {author} {\bibfnamefont
  {A.}~\bibnamefont {Steinhoff}}, \bibinfo {author} {\bibfnamefont
  {B.}~\bibnamefont {Pano}}, \bibinfo {author} {\bibfnamefont {J.}~\bibnamefont
  {Klein}}, \bibinfo {author} {\bibfnamefont {F.}~\bibnamefont {Jahnke}},
  \bibinfo {author} {\bibfnamefont {A.}~\bibnamefont {Holleitner}}, \ and\
  \bibinfo {author} {\bibfnamefont {U.}~\bibnamefont {Wurstbauer}},\ }\href
  {\doibase 10.1021/acs.nanolett.7b01304} {\bibfield  {journal} {\bibinfo
  {journal} {Nano Letters}\ }\textbf {\bibinfo {volume} {17}},\ \bibinfo
  {pages} {5229} (\bibinfo {year} {2017})}\BibitemShut {NoStop}%
\bibitem [{\citenamefont {Baranowski}\ \emph {et~al.}(2017)\citenamefont
  {Baranowski}, \citenamefont {Surrente}, \citenamefont {Klopotowski},
  \citenamefont {Urban}, \citenamefont {Zhang}, \citenamefont {Maude},
  \citenamefont {Wiwatowski}, \citenamefont {Mackowski}, \citenamefont {Kung},
  \citenamefont {Dumcenco}, \citenamefont {Kis},\ and\ \citenamefont
  {Plochocka}}]{baranowski_probing_2017}%
  \BibitemOpen
  \bibfield  {author} {\bibinfo {author} {\bibfnamefont {M.}~\bibnamefont
  {Baranowski}}, \bibinfo {author} {\bibfnamefont {A.}~\bibnamefont
  {Surrente}}, \bibinfo {author} {\bibfnamefont {L.}~\bibnamefont
  {Klopotowski}}, \bibinfo {author} {\bibfnamefont {J.~M.}\ \bibnamefont
  {Urban}}, \bibinfo {author} {\bibfnamefont {N.}~\bibnamefont {Zhang}},
  \bibinfo {author} {\bibfnamefont {D.~K.}\ \bibnamefont {Maude}}, \bibinfo
  {author} {\bibfnamefont {K.}~\bibnamefont {Wiwatowski}}, \bibinfo {author}
  {\bibfnamefont {S.}~\bibnamefont {Mackowski}}, \bibinfo {author}
  {\bibfnamefont {Y.~C.}\ \bibnamefont {Kung}}, \bibinfo {author}
  {\bibfnamefont {D.}~\bibnamefont {Dumcenco}}, \bibinfo {author}
  {\bibfnamefont {A.}~\bibnamefont {Kis}}, \ and\ \bibinfo {author}
  {\bibfnamefont {P.}~\bibnamefont {Plochocka}},\ }\href {\doibase
  10.1021/acs.nanolett.7b03184} {\bibfield  {journal} {\bibinfo  {journal}
  {Nano Letters}\ }\textbf {\bibinfo {volume} {17}},\ \bibinfo {pages} {6360}
  (\bibinfo {year} {2017})}\BibitemShut {NoStop}%
\bibitem [{\citenamefont {Zhao}\ \emph {et~al.}(2015)\citenamefont {Zhao},
  \citenamefont {Yang}, \citenamefont {Wang}, \citenamefont {Guo},\ and\
  \citenamefont {Ji}}]{zhao_continuously_2015}%
  \BibitemOpen
  \bibfield  {author} {\bibinfo {author} {\bibfnamefont {Y.-H.}\ \bibnamefont
  {Zhao}}, \bibinfo {author} {\bibfnamefont {F.}~\bibnamefont {Yang}}, \bibinfo
  {author} {\bibfnamefont {J.}~\bibnamefont {Wang}}, \bibinfo {author}
  {\bibfnamefont {H.}~\bibnamefont {Guo}}, \ and\ \bibinfo {author}
  {\bibfnamefont {W.}~\bibnamefont {Ji}},\ }\href
  {http://dx.doi.org/10.1038/srep08356} {\bibfield  {journal} {\bibinfo
  {journal} {Scientific Reports}\ }\textbf {\bibinfo {volume} {5}},\ \bibinfo
  {pages} {8356} (\bibinfo {year} {2015})}\BibitemShut {NoStop}%
\bibitem [{\citenamefont {Lloyd}\ \emph {et~al.}(2016)\citenamefont {Lloyd},
  \citenamefont {Liu}, \citenamefont {Christopher}, \citenamefont {Cantley},
  \citenamefont {Wadehra}, \citenamefont {Kim}, \citenamefont {Goldberg},
  \citenamefont {Swan},\ and\ \citenamefont {Bunch}}]{Lloyd_Band_2016}%
  \BibitemOpen
  \bibfield  {author} {\bibinfo {author} {\bibfnamefont {D.}~\bibnamefont
  {Lloyd}}, \bibinfo {author} {\bibfnamefont {X.}~\bibnamefont {Liu}}, \bibinfo
  {author} {\bibfnamefont {J.~W.}\ \bibnamefont {Christopher}}, \bibinfo
  {author} {\bibfnamefont {L.}~\bibnamefont {Cantley}}, \bibinfo {author}
  {\bibfnamefont {A.}~\bibnamefont {Wadehra}}, \bibinfo {author} {\bibfnamefont
  {B.~L.}\ \bibnamefont {Kim}}, \bibinfo {author} {\bibfnamefont {B.~B.}\
  \bibnamefont {Goldberg}}, \bibinfo {author} {\bibfnamefont {A.~K.}\
  \bibnamefont {Swan}}, \ and\ \bibinfo {author} {\bibfnamefont {J.~S.}\
  \bibnamefont {Bunch}},\ }\href {\doibase 10.1021/acs.nanolett.6b02615}
  {\bibfield  {journal} {\bibinfo  {journal} {Nano Letters}\ }\textbf {\bibinfo
  {volume} {16}},\ \bibinfo {pages} {5836} (\bibinfo {year} {2016})},\ \bibinfo
  {note} {pMID: 27509768}\BibitemShut {NoStop}%
\end{thebibliography}%

\end{document}